\documentclass{aa}
\usepackage{graphicx}
\usepackage{epsfig}
\usepackage{graphics}

\newcommand{\be}{\begin{equation}}
\newcommand{\ee}{\end{equation}}
\begin{document}%
\title{Leptonic secondary emission in a hadronic microquasar model}

   \author{M. Orellana\inst{1,2,}\thanks{Fellow of CONICET, Argentina},
           P. Bordas\inst{3},
           V. Bosch-Ramon\inst{4},
           G.~E. Romero\inst{1,2,}\thanks{Member of CONICET, Argentina},
      \and J. M. Paredes\inst{3}           
           }

   \offprints{Mariana Orellana {\em morellana@carina.fcaglp.unlp.edu.ar}}
   \titlerunning{Leptonic secondary emission in an hadronic MQ model}

\authorrunning{M. Orellana et al.}

\institute{Facultad de Ciencias Astron\'omicas y Geof\'{\i}sicas, Universidad Nacional de La Plata, Paseo del Bosque, 1900 La Plata, Argentina \and Instituto Argentino de Radioastronom\'{\i}a, C.C.5, (1894) Villa Elisa, Buenos Aires, Argentina \and 
Departament d'Astronomia i Meteorologia, Universitat de Barcelona, Mart\'{\i} i Franqu\`es 1, 08028, Barcelona, Spain\and Max Planck Institut f\"ur Kernphysik, Saupfercheckweg 1, Heidelberg 69117, Germany}

\date{Accepted 04/09/2007}
 \abstract
{It has been proposed that the origin of the very high-energy photons emitted from high-mass X-ray binaries with jet-like features, so-called microquasars (MQs), is related to
hadronic interactions between relativistic protons in the jet and cold protons of the stellar wind. Leptonic secondary emission should be
calculated in a complete hadronic model that include the effects of pairs from charged pion decays inside the jets and the emission from pairs generated by gamma-ray absorption in the photosphere of the system.}
{We aim at predicting the broadband spectrum from a general hadronic microquasar model, taking into account the emission from secondaries created by charged pion decay inside the jet.
}
{The particle energy distribution for secondary leptons injected along the jets is consistently derived taking the energy losses into account. 
We also compute the spectral energy distribution resulting from these leptons is calculated after assuming different values of the magnetic field inside the jets. The spectrum of the gamma-rays produced  by neutral pion-decay and processed by electromagnetic cascades under the stellar photon field. }
{We show that the secondary emission can dominate the spectral energy distribution at low energies ($\sim 1$ MeV). At high energies, the production
spectrum can be significantly distorted by the effect of electromagnetic cascades. These effects are phase-dependent, and some variability modulated by
the orbital period is predicted.}
{}

   \maketitle
%
\section{Introduction}

Many massive binary systems contain the final product of massive stellar evolution, a black hole or a neutron
star. The mass transfer from the less evolved star onto the compact companion forms an accretion disk and generates X-ray emission. Some of these X-ray binaries also display collimated outflows of relativistic particles (jets) resembling scaled-down versions of the extragalactic quasars. It is widely accepted that these jets are powered 
by accretion processes and/or the rotational energy of the compact object. Their collimation is thought to be related to the magnetic field structure in the surroundings of the compact object. 

Recently, microquasars (MQs) have been detected by ground-based Cherenkov telescopes at energies higher than 200 GeV
(Aharonian et al. 2005; Albert et al. 2006: Albert et al. 2007). These detections have confirmed MQs as    
sites of effective acceleration of particles up to ultra-relativistic energies. The jets are the best candidates for generating the very high-energy radiation since they also present extended non thermal radio emission  (see, e.g., Rib\'o 2005) and/or X-ray emission (see, e.g. Corbel et~al. 2002), evidencing the presence of a relativistic population of particles. Orbital modulation observed in some MQs at very different wavelengths and morphological
changes reveal a complex phenomenology, still not understood, taking place in these gamma-ray binaries (e.g. LS~I~+61 303; Dhawan et~al. 2006\footnote{We note that these authors have proposed
LS~I+61~303 as a pulsar binary system because the orbital variation of the extended radio  emission
morphology. For a critical assessment, see Romero et al. (2007).}).
For an updated review of the theoretical aspects concerning MQs, see Bosch-Ramon
(2007).

The nature of the matter outflowing in the jets of MQs is still an issue of debate, although  the presence of ions has been inferred from the detection of X-ray  iron lines in the case of SS~433 (Kotani et al. 1994; Migliari et al. 2002); hence, MQs jet models attempting to explain the
broad-band spectral features of these sources have been centered in two different approaches. For a treatment considering the leptonic nature of the radiative processes that produce the primary emission, see e.g. Bosch-Ramon et al. (2006a). On the other hand, the so-called hadronic models, until now proposed only for high-mass systems, have predicted high-energy emission from the interaction between protons accelerated in the jet and cold protons from the stellar wind of the primary star. The inelastic proton-proton collisions generate neutral pions that decay thereby producing the gamma-ray photons (see, e.g., Romero et al. 2003; Orellana \& Romero 2007). 

In hadronic models, the $\gamma$-ray emission should be accompanied by a flux of
high-energy neutrinos emerging from the decay of $\pi^\pm$. Their detection should be the unambiguous
confirmation of the hadronic interactions (Aharonian et al. 2006). Also, since hadrons suffer energy losses to a lesser extent than leptons, the high-energy cut-off of the spectrum in hadronic models can be expected to occur at higher energies than in pure leptonic models (Orellana \& Romero 2007). If there are other
features that can be tested at lower energies to support the hadronic model over
the pure leptonic one, they should be looked for in the secondary particle injection expected in such a scenario.

We explore in this work the emission of leptonic origin that is unavoidably produced in a hadronic scenario. There are two populations of secondary leptons that emerge in this context. On the one hand, relativistic leptons are injected along the jet by the
charged pion decay channel. They are expected to be confined in the jet by the magnetic field and exposed
to adiabatic and radiative energy losses of the same kind as in primary leptonic models. Regarding the radiative losses, we have
accounted here for synchrotron energy losses under the magnetic field present in the jet and inverse Compton (IC) interactions with photons coming from the radiation field of the companion star.
On the other hand, the substantial absorption of high-energy gamma-rays, mainly of neutral pion decay origin, in the photon field of the donor star injects energetic $e^{\pm}$ pairs in the photosphere of the system. These leptons scatter stellar photons to high energies through IC interactions. The gamma-rays can be absorbed again, and an electromagnetic cascade develops in this way. 

The steady-state pair and photon distributions resulting from the cascade reprocessing of the primary radiation can be estimated through numerical methods. We have performed Monte Carlo simulations of the electromagnetic cascade development along the line of sight. These one-dimensional cascades are induced by the primary hadronic emission when isotropization of very high-energy particles is not efficient and the magnetic field is rather weak, ensuring that the leptonic energy losses are IC-dominated. Variability of the outgoing spectra due to changes in the geometry of the IC scattering and photon-photon absorption along the orbit, in which the inclination of the system plays an important role, can be determined. In what follows we outline the model and present our calculations and results.  

\section{The general picture}
The general picture treated in this work is the same as presented by Romero et al. (2003). The binary system is formed by a high-mass star and a compact object, here assumed to be a black hole. The star blows a radiatively driven spherical wind that is partially accreted by the black hole and that developes an accretion disk. The mass accretion rate is assumed constant along the orbit, adopted here to be circular. Some of the accreted matter and energy is redirected into the outflowing jets. Hence, the total power of the jets is related to the total accretion energy budget.

\begin{figure*}
\centering \psfig{figure=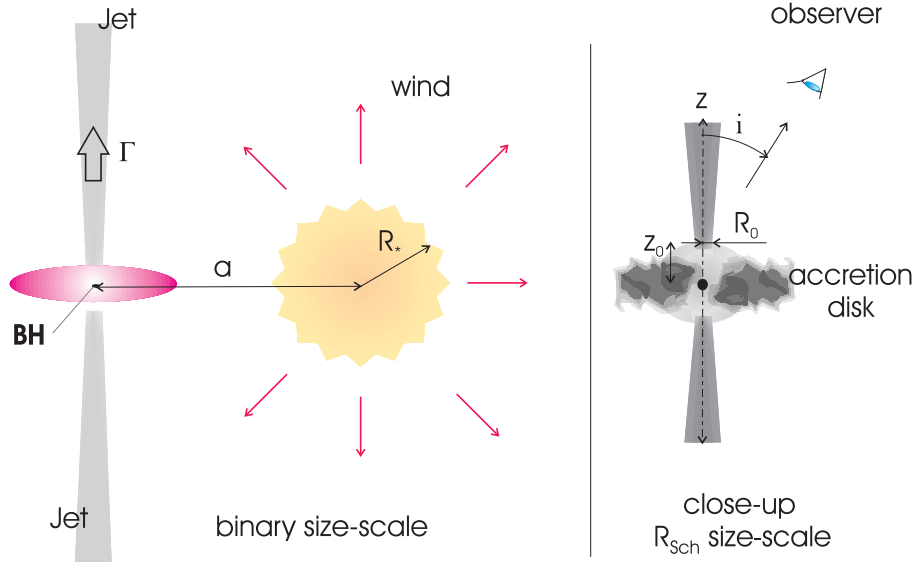,width=12cm}
\caption{Sketch of the configuration discussed in this paper. The main parameters are indicated. On the right a zoom-in of the region close to the black hole is shown.} 
\label{sketch}
\end{figure*}

We assume a continuous jet, as observed in the stable configuration of a MQ in a low-hard X-ray state (e.g. Fender et al. 2004). The launching mechanism for the jet is not treated here, and for simplicity, the jets are considered perpendicular to the orbital plane. These jets move away from the injection point at distance $z_0$ above and below the orbital plane and expand laterally as a cone with a jet semi-opening angle $\sim 5^{\circ}$. They propagate with a mildly relativistic velocity characterized by a bulk Lorentz factor $\Gamma=1.5$. Figure \ref{sketch} shows a sketch of the configuration.

The kinetic power of relativistic protons in each jet, $L_p^{\rm rel}$, is taken as a constant fraction ($\eta$) of the Eddington luminosity of the compact object. Two cases were considered, labeled as Model A for $\eta=0.005$ , i.e. $L_p^{\rm rel}=6\times 10^{36}$~erg~s$^{-1}$, and Model B for $\eta=0.0005$, i.e. $L_p^{\rm rel}=6\times 10^{35}$~erg~$s^{-1}$. We note that recently a low surface brightness arc of radio emission was discovered around the microquasar Cygnus X-1 and interpreted in terms of a shocked compressed shell of emitting gas inflated by the jets (Gallo et al. 2005). This discovery  led to the suggestion that the kinetic power of MQ jets can be as high as the bolometric X-ray luminosity of the system. For a system like Cygnus X-1, $L_X^{\rm bol}\sim 10^{37}$ erg s$^{-1}$; therefore, the values assumed above are consistent with this constraint.

We assume that the wind ions penetrate the jet laterally, resulting in an effective production of
charged and neutral pions through $pp$ collisions. The main production region has roughly the size of the binary system, since farther away the wind dilutes and the number of interactions decreases rapidly. A penetration factor of 0.3 has been introduced to take the mixing between the jet and the wind material into account in a phenomenological way (see Romero et al. 2005). 

Concerning the wind that provides the proton targets, we consider a radiatively driven wind for simplicity. The radial dependence of the matter velocity is $v(r)\propto(1-R_*/r)$ (Lamers \& Cassinelli 1999). The
particle density of this proton-dominated gas is obtained from the continuity equation. Along the jet (coincident with the $z$ axis), it yields

\be
n(z)=\frac{\dot{M}_\star}{4 \pi m_p (z^2+a^2)}\left( 1-\frac{R_\star}{\sqrt{z^2+a^2}}\right)^{-1}.
\ee
The parameters adopted for the models are listed in Table \ref{tab}. Mostly, they correspond to a similar system to Cyg X-1: a high-mass binary with orbital period of 5.6 d, formed by an O9.7 Iab super-giant primary (Walborn 1973) and a black hole of $M_{\rm BH}=10$ M$_\odot$. The inclination of the orbital plane is taken to be $i=30^\circ$. The orbit is circular but its orientation induces periodic changes in the optical depth for photons produced close to the compact object. We follow the phase notation of Szotek \& Zdziarski (2006): $\phi=0$
at the superior conjunction of the X-ray source, when the compact object is behind the primary. 
 
\begin{table}
  \caption[]{Model parameters. The primes indicate quantities measured in the jet co-moving frame.}
  \label{tab}
  \begin{center}
  \begin{tabular}{p{0.5cm}lll}
  \hline\noalign{\smallskip}
\multicolumn{2}{c}{Parameter, description [units]}&values\\
  \hline\noalign{\smallskip}
$e$ & eccentricity & 0 \\
$a$& orbital semi-major axis [$R_{\star}$] & 3 \\
$i$& orbital inclination [$^\circ$] & 30 \\ 
$M_{\star}$&stellar mass [$M_{\odot}$] & 30 \\
$\dot{M}_\star$& stellar mass loss rate [$M_{\odot}$~yr$^{-1}$] & 3 $\times 10^{-6}$ \\
$v_\infty$&terminal velocity of the wind [cm s$^{-1}$] & $2\times 10^8$\\
$M_{\rm BH}$&compact object mass [$M_{\odot}$] & 10\\
$R_{\rm 0}$&initial radius of the jet [$R_{\rm Sch}^1$]& 5 \\
$z_0$&jet initial point [$R_{\rm Sch}$] & 50 \\
$\chi$&jet semi-opening angle tangent & 0.1 \\
$\alpha$&proton power-law index & 2 \\
$\gamma\prime _p^{\rm min}$&proton minimum Lorentz factor& 2 \\
$\gamma\prime _p^{\rm max}$&proton maximum Lorentz factor$^\dagger$& $10^5$\\
$\Gamma$&macroscopic Lorentz factor & 1.5\\
$R_{\star}$&stellar radius [$R_{\odot}$] & 15  \\
$L_{\star}$&stellar bolometric luminosity [erg~s$^{-1}$] & 6 $\times 10^{38}$  \\
$T_{\star}$&stellar surface temperature [K] & $3\times10^4$ \\
$f_p$&wind-jet penetration factor & 0.3\\
\noalign{\smallskip}
\hline \noalign{\smallskip}\multicolumn{3}{l} {$^1$$R_{\rm Sch}=2GM_{\rm BH}/c^2$, $^\dagger$$E _p^{\rm max}\sim 100$~TeV}\cr
  \end{tabular}
  \end{center}
\end{table}

The jets are assumed to contain a randomly oriented magnetic field.
We consider that, because of adiabatic expansion, the magnetic field follows $B_{\rm
jet}\sim B_0(z_0/z)$, where $z_0=50$~$R_{\rm Sch}$. At the base of the jet we can constrain the value of the magnetic field by its equipartition value, that is, $B_0^{\rm eq}\sim \sqrt{8\pi U_p^{\rm rel}}$, with the energy density $U_p^{\rm rel}=L_p^{\rm
rel}/(\pi R_0^2 c)$. Here, $R_0$ is the initial radius of the jet, i.e. at $z_0$. Then, $B_0^{\rm eq}\sim 10^6$ G and $B_0^{\rm eq}\sim 10^5$ G result for Models A and B, respectively. When emission from secondaries in the jet is computed (see Section \ref{sec}), we adopt a magnetic field corresponding to $10 \%$ of the equipartition values, incorporating an efficiency factor, as in models with cold-matter dominated jets (Bosch-Ramon et al., 2006a).

\section{Hadronic primary emission}

We outline here the basics of the hadronic model for primary $\gamma$-ray production. The reader is referred to Romero et al. (2003 and 2005) for comprehensive descriptions.

The particle spectrum of relativistic protons is assumed to be a power law in the jet reference frame with the canonical index ($\alpha=2$) from shock diffusive acceleration. The number density of particles entering the jet per unit of time can be determined as in Romero et~al. (2003). The conservation of the number of particles makes the proton flux depend on the distance to the compact object as $J_p(z)\propto (z_0/z)^2$. The Lorentz transformation that relates the co-moving 
jet frame with the observer's frame introduces an angular dependence in the proton injected 
flux (Purmohammad \& Samimi, 2001). Specifically, we have
$$J_{\!p}(\!E_p,\theta)\!=\!\frac{c K_0}{4 \pi} \!\!\left(\frac{z_0}{z}\right)^
{\!2} \!\frac{\Gamma^{-\alpha+1}\! \!\left(\!E_p\!-\!\beta_{\rm b}
\sqrt{E_p^2-m_p^2c^4} \cos \theta\right)^{\!\!-\alpha}}{\left[\sin ^2
\theta +\! \Gamma^2 \!\left( \!\cos \theta - \!\frac{\beta_{\rm b}
E_p}{\sqrt{E_p^2-m_p^2 c^4}}\!\right)^{\!\!2}\right]^{\! 1/2}} 
$$
\noindent where $\Gamma$ is the jet Lorentz factor, and $\beta_{\rm b}$
 the corresponding velocity in units of $c$. The constant $K_0$ normalizes the energy distribution of the proton flux in the jet's co-moving frame. The zenith angle $\theta$ is measured from the $z$ axis. 
The gamma-ray photons are beamed along the proton direction, reproducing their angular distribution. 
The maximum value of $E_p$ obtained by balancing the acceleration rate with the rate of energy loss by synchrotron, $pp$ and $p\gamma$ 
interactions exceeds the energy constraints imposed by the equality between the maximum particle gyroradius and the size of the accelerator, the latter taken as 0.1 times the jet radius 
at each height ($R_{\rm jet}(z)=\chi z$). 
In a model where the dependence of the magnetic field strength on $z$ is inverse to that 
of $R_{\rm jet}(z)$, $E_p^{\rm max}$ is constant along the jet, i.e. $E_p^{\rm max}=0.1eB_0R_0\sim100$~TeV. This value is used to impose an exponential cut-off on the above proton-flux.

In our model, the generation of primary gamma-ray emission is not affected by the circular orbital motion of the compact object. With the parameters of Table~\ref{tab} we  have computed the spectral energy distribution (SED) of gamma-ray photons resulting from neutral pion decays. The $\delta$-functional  approximation is used to estimate the production rate of $\pi^0$-mesons\footnote{For a comparison between the the $\delta$-functional approximation and other parametrizations of the $\gamma$-ray and $e^\pm$ emissivities, see Domingo-Santamar\'{\i}a \& Torres (2005).}. Kelner et al. (2006) provide a recent and accurate approximation  for the inelastic cross-section: $\sigma_{\rm inel}(E_p)=(34.3+1.88\,L+0.25\,L^2)\times\left[1-\left(E_{\rm th}/E_p\right)^{4}\right]^2$ mb,  
where $E_{\rm th}=1.22$ GeV is the threshold energy of production of $\pi^0_{}$-mesons, and
$L=\ln(E_p/1\,{\rm TeV})$.

\begin{figure}
\centering \psfig{figure=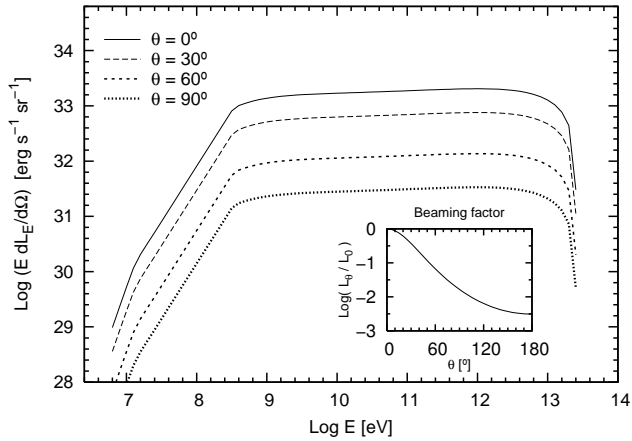,width=10.0cm,angle=-90}
\caption{Gamma-ray luminosity per unit of solid angle produced by neutral pion decays in Model A ($L_p^{\rm rel}=6\times 10^{36}$ erg s$^{-1}$). In decreasing order, the curves correspond to $\theta$= 0, 30, 60, and 90 degrees. The beaming factor is shown in the small panel.} 
\label{L_o}
\end{figure}

In Fig.~\ref{L_o} we show the hadronic gamma-ray luminosity generated along the jet (integrated along $z$) for
different $\theta$ directions. The beaming factor is defined by the ratio $L(\theta)/L(\theta=0)$ for energies
greater than 10 GeV, where the angular dependence of the emerging flux is almost energy independent.

\section{Secondary emission inside jets}\label{sec}

The expressions given by Kelner et al. (2006) are used to compute the leptonic energy distribution of $e^\pm$
injected along the jet by the charged pion decays. The resulting injection rate (leptons per units of energy, volume, and time) can be expressed in terms of the variable $x=E_{e^\pm}/E_p$ as
\be
q_{e}(E_{e})=c\,n_{H}^{}\int\limits_0^1\! \sigma_{\rm inel}(E_{e}/x)\,
J_p(E_{e}/x)\,F_{e}(x,\,E_{e}/x)\,\frac{dx}{x}.
\label{Phi g2}
\ee
Here, $F_{e}(x,\,E_{e}/x)$ is
the spectrum of electrons from the $\pi\to\mu\,\nu_\mu$ decay, and it is described by 
\be
\label{LEP1}
F_{e}(x,\,E_p)=C_{1}(E_{p})\,\frac{(1+C_{3}(E_{p})(\ln x)^2)^3}{x+0.3x^{(1-C_{2}(E_{p}))}}\,(-\ln(x))^5\,,
\ee
where 
\begin{eqnarray}
C_{1}(E_{p})&=&\frac1{69.5+2.65\,L+0.3\,L^2}\,,\\[4pt]
C_{2}(E_{p})&=&\frac1{(0.201+0.062\,L+0.00042\,L^2)^{1/4}}\,,\\[4pt]
C_{3}(E_{p})&=&\frac{0.279+0.141\,L+0.0172\,L^2}{0.3+(2.3+L)^2}\,,
\end{eqnarray}
and $L=\ln(E_p/1\,{\rm TeV})$.

For the injected proton flux (see Section 3), we obtain $q_e(E_e)\sim 3\times 10^{27} (E_e/{\rm eV})^{-1.92}$ leptons erg$^{-1}$ s$^{-1}$ cm$^{-3}$, at the base of the jet, in the observer's reference frame for Model B (an order of magnitude higher for Model A).
Once produced, we assume that the linear momentum of the leptons loses its angular dependence, erased by the random nature of the magnetic field. In the calculations, we integrate the $\theta$ dependence of the differential rate of production of $e^\pm$ and treat the population as isotropic in the jet reference frame. 
The leptons injected along the jet suffer from radiative cooling due to synchrotron and IC processes, as well as adiabatic expansion losses. The evolution of the particle energy distribution is studied by dividing the jet into slices of suitable size in order to consider the physical conditions as homogeneous within each of them. Afterward, to obtain the distribution evolution with $z$, we solve the continuity differential equation. Note that different evolved
populations sum up at each height, since injection occurs all along the jet. The expressions used are given in Bordas et al. (2007). 

The secondary synchrotron emission is calculated in the presence of the magnetic field, which is determined by its value at the base of the jet $B_0$.
We compute the emission for $B_0=10^5$~G (Model A) and $B_0=10^4$~G (Model B). 
We assume the same lepton maximum energy, since for simplicity  $E_p^{\rm max}$ is fixed to 100~TeV
for both values of the magnetic field. For the specific emission and absorption coefficients, we used the expressions given by Pacholczyk (1970). We then computed the spectral energy distributions and transformed them into the observer frame. 

As a first-order approach, we assumed that the IC takes place for any lepton energy in the Thomson regime and under an isotropic photon field. Actually,
the Thomson approximation is not valid for particles with energies above $\ga 10$--$100$~GeV, i.e. at the highest energies of the secondary leptons produced in this scenario. Above these energies, the IC interactions occur in the Klein Nishina (KN) regime. This fact does not affect the radiated synchrotron spectrum significantly, since it is mainly produced in the inner regions of the jet, close to the compact object, where the synchrotron process is the main cooling mechanism; i.e. the particle spectrum is not noticeably affected by KN cooling there. 
The KN effect would otherwise affect the highest energies of the IC spectrum, but this energy range of the spectrum is dominated by neutral pion decay $\gamma$-ray photons (see next section). 

\begin{figure}
\centering \psfig{figure=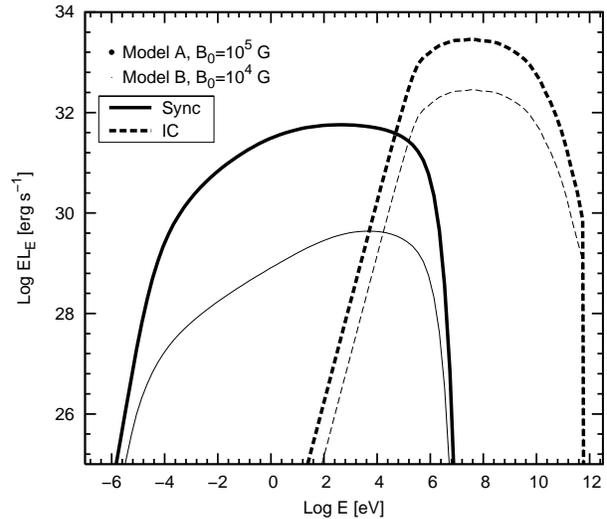, width=7.0cm,angle=-90}
\caption{Synchrotron and inverse Compton emission from the leptons produced inside the jet by the charged pions' decay. The inverse Compton interactions are calculated in the Thomson regime.}\label{pol}
\end{figure}

In Figure \ref{pol}, we show the synchrotron and IC spectra.
These calculations allow us to estimate the power going to this radiative channel in order to compare it with other components like IC emission from electromagnetic cascades or synchrotron emission from the secondaries inside the jet. 
The set of explored parameters illustrate the impact of the IC/synchrotron energy loss balance in the secondary broadband spectrum. 
It is seen that the synchrotron luminosity is at most $\sim 10^{32}$~erg~s$^{-1}$, and well below the IC emission for the both values of the adopted magnetic field. 

\section{IC cascades along the line of sight}

\begin{figure}
\centering \psfig{figure=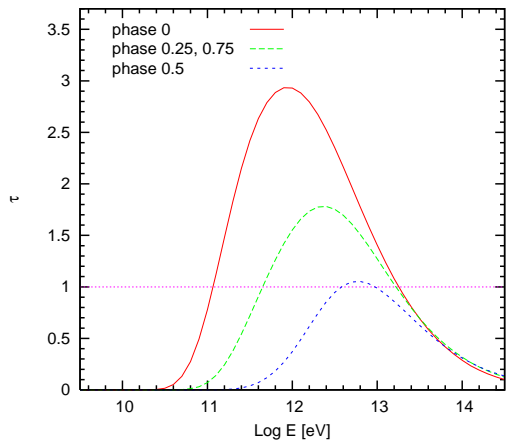, width=6.0cm,angle=-90}
\caption{Optical depth for gamma-ray photons injected at $z=a/2$. The absorber photon field generated by the donor star is anisotropic, considered here as a black body emitter. Each curve corresponds to a fixed orbital phase, with $\phi=0$ at the superior conjunction of the X-ray source, when the compact object is behind the primary.}\label{tau}
\end{figure}

The radiation field of early-type stars provides a suitable target for the absorption of gamma-ray photons. In
Fig.~\ref{tau} we show the optical depth as a function of the gamma-ray energy for the injection at height
$z=a/2$.  Hereafter we consider that as a representative point for the injected primary emission. The plotted
optical depths were calculated as in Dubus (2006). The absorption probability is modulated on the orbital period
of the binary as a result of the phase dependence of the geometric path (see the sketch in Szostek \&
Zdziarski 2006).

Energetic electron-positron pairs are materialized by photon-photon interactions. These leptons, in turn,
boost the stellar photons to high-energies via IC scattering.  The $\gamma$-ray absorption and production
mechanisms can proceed very fast, resulting in the development of an electromagnetic cascade (Bednarek 1997,
2000, 2006; Aharonian et~al. 2006). As a result, the energy of the original photons is distributed between
a certain number of secondary particles and photons with lower energy. 

We computed the $\gamma$-ray spectra formed in cascades traversing the anisotropic stellar
radiation field. Other radiative fields (like the accretion disk field) are less important in
the present context. We implemented a Monte Carlo simulation code based on the scheme outlined by Protheroe
(1986) and Protheroe et al. (1992). A description of the treatment is given in Orellana et al. (2006).

The injected flux that initiates the cascades is obtained by fitting the gamma-ray spectrum for $i=\theta=30$
degrees. For Model A, we find\footnote{The isotropic luminosity is quoted here, being defined $L(E)= 4 \pi dL(E)/d\Omega$.}
$\log L (E_\gamma)\,{\rm [erg\,s^{-1}]}\sim 32.27+0.06 \log E_\gamma\,{\rm  [eV]}$  for 1 GeV $\le E_\gamma \le$
10 TeV.  The absorbing radiation field has a black-body spectrum. It comes from the O-type star, characterized
by $T_{\rm eff}\sim 30000$ K and $R_\star\sim10^{12}$ cm. 

For simplicity, any effects of the presence of a magnetic field are disregarded in these simulations. However, we note that the cooling rates by synchrotron and IC processes are equal at a small value of the magnetic field, namely $B\sim 0.1-1$ G for electron energies $\sim 1$ TeV, deep in the
KN regime. Hence, the value assumed for the magnetic field is crucial for the development of the cascade. 
The geometry and radial dependence of the stellar magnetic field close to the star has been derived by several authors (e.g., Eichler \& Usov 1993) under different simplifying assumptions. The main parameter for $B(r)$ is the value of magnetic field at the stellar surface, which is not well known, but it is thought that it could reach $\sim 10^3$ G at most. Then, the field could be in the range
$10^{-2}-10$ G in a region at a few stellar radii around a typical O star\footnote{These values of $B$ are taken from Benaglia \& Romero (2003) for colliding wind regions of WR+OB systems. Concerning the
stellar magnetic field the physical context is the same than the one treated here.}. 

As the cascade evolves, departing from the injection point toward the observer, it will certainly reach regions where the stellar magnetic field has decayed; hence, there is still some room for our assumption (small or negligible $B$) to be valid. 
Effects of the magnetic field influencing the pure IC cascade by changing the lepton direction of propagation have been incorporated by Bednarek (1997 and subsequent works) to cascade simulations. 
We nevertheless assume that, at the lepton energies to which most of the cascade develops, particles radiate before isotropizing after their creation. 

\begin{figure}
\centering \psfig{figure=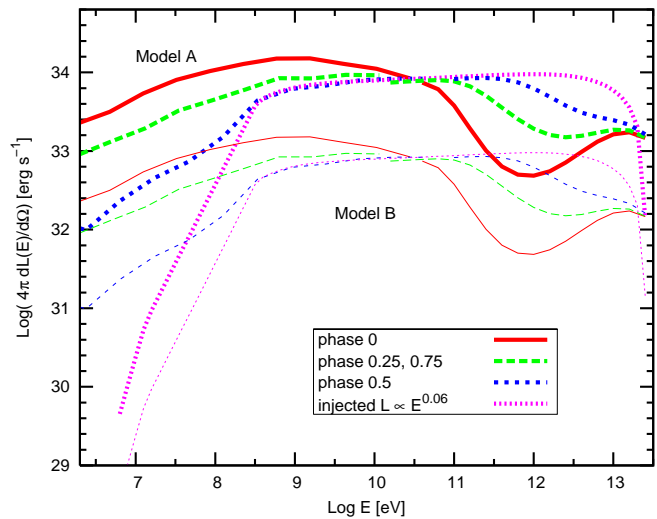, width=7.0cm,angle=-90}
\caption{Isotropic luminosity of cascade spectra for relevant orbital phases. 
The injected (hadronic) primary gamma-ray power is indicated. Thick lines correspond to Model A and the thin ones to Model B. For the orbital phases, $\phi=0$ when the compact object is behind the star, $\phi=0.5$ when it is in the front.
}\label{cascadaIC}
\end{figure}

The farthest distance from the source up to which we follow the cascades is $\sim 10$ times the orbital separation. Photons that
pass beyond that distance are stored to form the outgoing spectra.  At distances greater than the orbital separation, the cascade efficiency is reduced by the dilution of the stellar radiation field.  The spectra are obtained by constructing histograms $f(E)$, sorting photons into bins of width $\Delta (\log E)=0.5$. Some emerging spectra are shown in Fig.~\ref{cascadaIC}. 


The injected primary power is clearly re-distributed to lower energies. At energies higher than $\sim 10$ GeV, the spectra do not follow a simple power-law, but present a more complex structure, much softer than the injected spectrum. 
At energies below $\sim 100$ MeV, the cascade effect is essential for raising the luminosity, and the phase dependence is stronger than at higher energies. The results point to a time dependence anti-correlated for $\sim$GeV signal compared with that at $\sim$ TeV.

\section{Discussion and summary}
We have considered the simplified geometry of a circular orbit and a spherical wind for a high-mass MQ in order to illustrate the secondary emission produced in a hadronic model. We considered a power-law distribution with a spectral index $\alpha=2$ for the relativistic proton flux injected at the base of the jet. 
The gamma-ray and leptonic emissivities from $\pi$-meson decays were computed using the latest available analytic approximations for the inelastic $pp$ cross-section (Kelner et al. 2006). Relavistic leptons that are produced inside the jet via charged pion decays radiate subsequently via synchrotron and IC processes. 

The primary hadronic emission is found to be strongly affected by the pure IC cascade development in the photosphere of the primary star. We note that the synchrotron dominance of the leptonic energy losses can prevent the pure IC cascade from developing. In our treatment, we restrict ourselves to a low magnetic field in the region where the cascade occurs. In the case of a strong magnetic field, cooling via synchrotron radiation
 of the leptons injected by $\gamma-\gamma$ absorption will suppress the 
cascade development (e.g. Khangulyan et al. 2007).

The treatment implemented for the simulation of the electromagnetic cascades ignores the effect of any other process than IC interactions in the direction to the observer. This is a good approximation for when particles radiate before isotropize.
Efficient isotropization of $\sim 0.1$--$10$~TeV particles would require diffusion coefficients close to the Bohm one and magnetic field values $\ga 1$~G. 
In addition, Doppler boosting can also favor the jet radiation to dominate. Nevertheless, low-energy particles absorbed in different directions from that of the observer can present lower energy emission re-directed toward the observer since their direction of motion could indeed be randomized. Approximate estimates of the emitted radiation by leptons confined relatively close to the binary system 
are given in Bosch-Ramon (2007).

The cascade reprocessing can lead to significant emission 
at energies well below those originally injected (with primary energy $E_\gamma > 1$ GeV, due to the pion-creation threshold). The total radiation produced inside and outside the jet by charged pion and cascade secondaries is similar to that produced at gamma rays by the neutral pion decay. Detectability under the parameters assumed here can be discussed at different bands. 

At radio frequencies, the presented spectra are significantly hard, yielding 
 a flux $\sim$mJy at distances of a few kpc, which represents $\sim 10 \%$
of the fluxes found in MQs in the low-hard state, i.e. $\sim 15$~mJy (for Cygnus~X-1, see, e.g., Pooley et~al. 1999). If this radio emission produced by secondaries could be distinguished when observed at very high resolution, it would appear morphologically as a jet-like source surrounded by a radio halo, because of the highly beamed relativistic leptons generated by charged pion decays, on the one hand, and those generated by pair creation in the stellar photon field on the other. From these points, it seems that a primary leptonic component would be needed to explain the observed radio fluxes. Nevertheless, electrons have lifetimes that are long enough to reach regions of more suitable conditions to emit in the radio band, and the secondaries created in the jet could also be reaccelerated once they have lost their energy. Thus, it cannot be ruled out that extended radio emission is in fact produced by secondaries, in which case a primary electron population might not be necessary for explaining the non thermal spectrum observed in these systems. Such a case is outside of the assumptions made along this work, but could be interesting to explore in the future. 

At soft X-rays, synchrotron emission from secondaries could be detectable by the current X-ray instruments, although low-hard state MQs used to present significant X-ray emission produced in the accretion disk and the corona, which could mask this non thermal contribution to some extent. 

At hard X-rays, it is still not clear whether the radiation observed in some objects has a thermal or non thermal origin (see, e.g. Markoff et~al. 2005; Maccarone 2005; and also, Bosch-Ramon et~al. 2006c). In any case, the luminosities at hard X-rays obtained here by summing different contributions would not suffice for explaining those observed in some MQs during the low-hard state. 

The radiation in the MeV range seems to be non negligible, reaching $\sim 10^{33}$~erg~s$^{-1}$, although it still appears too low to be detected on a reasonable time span for the present instruments. At high and very high-energy gamma rays, GLAST and modern Cherenkov telescopes like
HESS or MAGIC could detect the source, which could present a relatively soft spectrum due to electromagnetic cascading. In fact, while finalizing the research for this paper, the detection of Cygnus X-1 by the MAGIC collaboration was announced (Albert et al. 2007). Throughout this work, it was not our purpose to explain the detailed behavior of this specific source; instead, we used some of the parameters of Cyg X-1 to apply the general treatment to this very interesting example. However, it is remarkable that the source has been detected by MAGIC around phase 0.9, when photon-photon absorption is high. Actually, it may be related to some additional variability of the production mechanism not accounted for here but linked to this phase (e.g. affected by perturbations in the stellar wind material; e.g. Feng \& Cui 2002), although the investigation of such a possibility is left for future work.

Even from the simple model considered here, we have been able to discuss variable emission in some energy bands. Many specific factors should lead to more complex variations (see e.g.  Bosch-Ramon et al. 2006a).
For instance, realistic modeling of the variability would include a variable accretion rate (e.g. induced through the variable nature of the outflows of the donor star, or simply through an eccentric orbit) and possible changes through the misalignment between the jet and the orbital axis (i.e. a precessing jet, see Kaufman
Bernad\'o et al. 2002). 
In such a more complex scenario, it would also be worthwhile study the evolution along the orbit of the IC hard X-rays and soft gamma rays due to the angular dependence in the cross section of this mechanism (see Khangulyan et al. 2007 and references therein).

\begin{acknowledgements}
We thank Dr. D. Khangulyan for useful comments.
M.O. thanks Universitat of Barcelona for support and hospitality while visiting this institution.
This research was supported by CONICET (PIP 5375) and the Argentine agency ANPCyT through Grant PICT 03-13291 BID 1728/OC-AR. P.B. was supported by the DGI of MEC (Spain) under fellowship BES-2005-7234. 
V.B-R. thanks the Max-Planck-Institut f\"ur Kernphysik for its support and kind hospitality. V.B-R. gratefully acknowledges support from the Alexander von Humboldt Foundation.
P.B., V.B-R., and J.M.P 
acknowledge support by the DGI of MEC under grant
AYA2004-07171-C02-01, as well as partial support by the European Regional Development Fund (ERDF/FEDER).
\end{acknowledgements}
{}
\end{document}